\documentclass[runningheads]{llncs}

\usepackage{graphicx}
\usepackage{amssymb,amsmath,color}
\usepackage[breaklinks=true]{hyperref}

\usepackage[anythingbreaks]{breakurl}

\newcommand{\beq}{\begin{equation}}
\newcommand{\eeq}{\end{equation}}
\newcommand{\ben}{\begin{enumerate}}
\newcommand{\een}{\end{enumerate}}

\title{Risk-Limiting Audits by Stratified Union-Intersection Tests of Elections (SUITE)}

\author{
   Kellie Ottoboni\inst{1} \and
   Philip B.~Stark\inst{1} \and
   Mark Lindeman\inst{2} \and
   Neal McBurnett 
}
\authorrunning{K.~Ottoboni et al.}
\titlerunning{SUITE Risk-Limiting Audits}

\institute{
Department of Statistics, University of California, Berkeley, CA, USA \and
Verified Voting Foundation, Philadelphia, PA, USA}

\date{Version: \today}

\begin{document}
\maketitle

\begin{abstract}
Risk-limiting audits (RLAs) offer a statistical guarantee: if a full manual tally of the paper ballots would show that the reported election outcome is wrong, an RLA has a known minimum chance of leading to a full manual tally.
RLAs generally rely on random samples.
Stratified sampling---partitioning the population of ballots into disjoint
strata and sampling independently from the strata---may simplify logistics or increase efficiency compared
to simpler sampling designs, but makes risk calculations harder.
We present SUITE, a new method for conducting RLAs using stratified samples.
SUITE considers all possible partitions of outcome-changing error across strata.
For each partition, it combines $P$-values from stratum-level tests into a combined
$P$-value; there is no restriction on the tests used in different strata.
SUITE maximizes the combined $P$-value over all partitions of outcome-changing error. 
The audit can stop if that maximum is less than the risk limit.
Voting systems in some Colorado counties (comprising 98.2\% of voters)
allow auditors to check how the system interpreted each ballot, 
which allows \emph{ballot-level comparison} RLAs.
Other counties use \emph{ballot polling}, which is less efficient.
Extant approaches to conducting an RLA of a statewide contest would require major changes to 
Colorado's procedures and software, or would sacrifice the efficiency of ballot-level comparison.
SUITE does not. 
It divides ballots into two strata: those cast in counties that can conduct ballot-level comparisons, and the rest.
Stratum-level $P$-values are found by methods derived here.
The resulting audit is substantially more efficient than statewide ballot polling.
SUITE is useful in any state with a mix of voting systems or that uses stratified sampling
for other reasons.
We provide an open-source reference implementation and exemplar calculations in Jupyter notebooks.

\end{abstract}

\noindent
\textbf{Acknowledgements.}
We are grateful to Ronald L.~Rivest and Steven N.~Evans for helpful conversations and suggestions.

\section{Introduction}
A risk-limiting audit (RLA) of an election contest is a procedure that
has a known minimum chance of leading to a full manual tally of the ballots if the electoral outcome  according to that tally would differ from the reported outcome.
\emph{Outcome} means the winner(s) (or, for instance, whether there is a runoff)---not the numerical vote totals. 
RLAs require a durable, voter-verifiable record of voter intent, such as paper ballots,
and they assume that this audit trail is sufficiently complete and accurate that a full hand
tally would show the true electoral outcome.
That assumption is not automatically satisfied: a \emph{compliance audit}
\cite{starkWagner12} 
is required to check whether the paper trail is trustworthy.

Current methods for risk-limiting audits are generally \emph{sequential hypothesis testing procedures}: they examine more ballots, or batches of ballots,
until either (i)~there is strong statistical evidence that a full hand tabulation would confirm the outcome,
or (ii)~the audit has led to a full hand tabulation, the result of which should become the official
result.

RLAs have been conducted in California, Colorado, Indiana, Ohio, Virginia, and Denmark, and are required by law in Colorado (CRS~1-7-515) and Rhode Island (SB~413A and HB~5704A).

The most efficient and transparent sampling design for risk-limiting audits selects individual ballots uniformly at random, with or without replacement~\cite{stark10c}.
Risk calculations for such samples can be made simple without sacrificing rigor~\cite{stark10d,lindemanStark12}.
However, to audit contests that cross jurisdictional boundaries then requires coordinating sampling in different counties, and may require different counties to use the lowest common denominator method for assessing risk from the sample, which would not take full advantage of the capabilities of some voting systems. 
For instance, any system that uses paper ballots as the official record can conduct \emph{ballot-polling} audits, while \emph{ballot-level comparison audits} require systems to generate \emph{cast vote records} that can be checked manually against a human reading of the paper~\cite{lindemanEtal12,lindemanStark12}. 
(These terms are described in Section~\ref{sec:combiningMethods}.) 

Stratified RLAs have been considered previously, primarily to conform with legacy audit laws under which counties draw audit samples independently of each other, but also to allow auditors to start the audit before all vote-by-mail or provisional ballots have been tallied, by sampling independently from ballots cast in person, by mail, and provisionally, as soon as subtotals for each group are available \cite{stark08a,higginsEtal11}.
However, extant methods address only a single approach to auditing, batch-level comparisons, and only a particular test statistic.

Here, we introduce SUITE, a more general approach to conducting RLAs using stratified samples.
SUITE is a twist on \emph{intersection-union} tests \cite{pesarinSalmaso10}, which represent the null hypothesis
as the intersection of a number of simpler hypotheses, and the alternative hypothesis as a union of their alternatives.
In contrast, here, the null is the union of simpler hypotheses, and the alternative is the intersection of their alternatives.
The approach involves finding the maximum $P$-value over a vector of nuisance parameters that describe the simple hypotheses, 
all allocations of tabulation error across strata for which a full count would find a different electoral outcome than was reported.
(A \emph{nuisance parameter} is a property of the population that is not of direct interest, but that affects the probability distribution of the data. 
\emph{Overstatement} is error that made the margin of one or more winners over one or more losers appear larger than it really was.
The total overstatement across strata determines whether the reported outcome is correct; 
the overstatements in individual strata are nuisance parameters that affect the distribution of the audit sample.)

The basic building block for the method is testing whether the overstatement error in a single stratum is greater than or equal to a quota.
Fisher's combining function is used to merge $P$-values for tests in different strata into a single $P$-value for the hypothesis that the overstatement in every stratum is greater than or equal to its quota.
If that hypothesis can be rejected for \emph{all} stratum-level quotas that could change the outcome---that is, if the maximum combined $P$-value is sufficiently small---the audit can stop.

It is not actually necessary to consider all possible quotas: the $P$-value involves a sum of monotonic functions, which allows us to find upper and lower bounds everywhere using only values on a discrete grid.
We present a numerical procedure, implemented in Python,
to find bounds on the maximum $P$-value when there are two strata.
The procedure can be generalized to more than two strata.

Section~\ref{sec:stratified} presents the new approach to stratified auditing.
Section~\ref{sec:combiningMethods} illustrates the method by solving a problem pertinent to Colorado:
combining ballot polling in one stratum with ballot-level comparisons in another.
This requires straightforward modifications to the mathematics behind ballot polling and ballot-level comparison to allow the overstatement to be compared to specified thresholds other than the overall contest margin; those modifications are described in Sections~\ref{sec:comparisonError} and~\ref{sec:ballotPollError}.
Section~\ref{sec:examples} gives numerical examples of simulated audits, using parameters intended to reflect how the procedure would work in Colorado.
We provide example software implementing the risk calculations for
our recommended approach in Python Jupyter notebooks.\footnote{%
 See \url{https://github.com/pbstark/CORLA18}.
}
Section~\ref{sec:discussion} gives recommendations and
considerations for implementation.

\section{Stratified audits} \label{sec:stratified}

\emph{Stratified sampling} involves partitioning a population
into non-overlapping groups and drawing independent random samples from those groups.
\cite{stark08a,higginsEtal11} developed RLAs based on comparing stratified samples of batches of ballots to hand counts of the votes in those batches: batch-level comparison RLAs, using a particular test statistic.
The method we develop here is more general and more flexible: it can be used with any test statistic, and test statistics in different strata need not be the same---which is key to combining audits of ballots cast using diverse voting technologies.

Here and below, we consider auditing a single plurality contest at a time, although the same sample can be used to audit more than one contest (and super-majority contests), and there are ways of combining audits of different contests into a single process \cite{stark09c,stark10d}.
We use terminology drawn from a number of papers, notably \cite{lindemanStark12}.

An \emph{overstatement error} is an error that caused the margin between \emph{any} reported
winner and \emph{any} reported loser to appear larger than it really was.
An \emph{understatement error} is an error that caused the margin between \emph{every} reported
winner and \emph{every} reported loser to appear to be smaller than it really was.
Overstatements cast doubt on outcomes; understatements do not, even though they are tabulation errors.

We use $w$ to denote a reported winner and $\ell$ to denote a reported loser.
The total number of reported votes for candidate~$w$ is $V_{w}$ 
and the total for candidate~$\ell$ is $V_{\ell}$.
Thus $V_{w} > V_{\ell}$, since $w$ is reported to have gotten more votes than $\ell$.

Let $V_{w\ell} \equiv V_{w} - V_{\ell} > 0$ denote the contest-wide margin (in votes) of 
$w$ over  $\ell$.
We have $S$ strata.
Let $V_{w\ell,s}$ denote the margin (in votes) of reported winner $w$ over reported loser $\ell$
in stratum $s$. 
Note that $V_{w\ell,s}$ might be negative in one stratum, but $\sum_{s=1}^S V_{w\ell,s} = V_{w\ell} > 0$.
Let $A_{w\ell}$ denote the margin (in votes) of reported winner $w$ over reported loser $\ell$ that  a full hand count would show: the \emph{actual} margin, in contrast to the \emph{reported} margin $V_{w\ell}$.
Reported winner $w$ really beat reported loser $\ell$ if and only if $A_{w\ell} > 0$.
Define $A_{w\ell,s}$ to be the actual margin (in votes) of $w$ over $\ell$ in stratum $s$.

Let $\omega_{w\ell,s} \equiv V_{w\ell,s} - A_{w\ell,s}$ be the \emph{overstatement}
of the margin of $w$ over $\ell$ in stratum $s$.
Reported winner $w$ really beat reported loser 
$\ell$ if and only if $\omega_{w\ell} \equiv \sum_s \omega_{w\ell,s} < V_{w\ell}$.

An RLA is a test of the hypothesis that the outcome is wrong, that is, that $w$ did not really beat $\ell$:
$\sum_s \omega_{w\ell, s} \ge V_{w\ell}$.
The null is true if and only if there exists \textit{some} $S$-tuple of real numbers $(\lambda_s)_{s=1}^S$ with $\sum_s \lambda_s = 1$ such that $\omega_{w\ell, s} \ge \lambda_s V_{w\ell}$ for all $s$.\footnote{%
  ``If'' is straightforward. For ``only if,'' suppose $\omega_{w\ell} \ge V_{w\ell}$. Set $\lambda_s = \frac{\omega_{w\ell, s}}{\sum_t \omega_{w\ell, t}}$. Then $\sum_s \lambda_s = 1$, and $\omega_{w\ell, s} = \lambda_s \omega_{w\ell} \ge \lambda_s V_{w\ell}$ for all $s$.
}
Thus if we can reject the conjunction hypothesis $\cap_s \{ \omega_{w\ell,s} \ge \lambda_s V_{w\ell} \}$ at significance level $\alpha$ for all $(\lambda_s)$ such that $\sum_s \lambda_s = 1$, we can stop the audit, and the risk limit will be $\alpha$.

\subsection{Fisher's combination method}

Fix $\mathbf{\lambda} \equiv (\lambda_s)_{s=1}^S$, with $\sum_s \lambda_s = 1$.
To test the conjunction hypothesis that stratum null hypotheses are true, that is, that $\omega_{w\ell,s} \ge \lambda_s V_{w\ell}$ for all $s$, we use Fisher's combining function.
Let $p_s(\lambda_s)$ be the $P$-value of the hypothesis $\omega_{w\ell,s} \ge \lambda_s V_{w\ell}$.
If the null hypothesis is true, then 
\beq \label{eq:fisher}
   \chi(\mathbf{\lambda}) = -2 \sum_{s=1}^S \ln p_s(\lambda_s)
\eeq
has a probability distribution that is dominated by the chi-square distribution with $2S$~degrees
of freedom.\footnote{%
   If the stratum-level tests had continuously distributed $P$-values, the distribution would be exactly
   chi-square with $2S$~degrees of freedom, but if any of the $P$-values has atoms when
   the null hypothesis is true, it is in general stochastically smaller.
   This follows from a coupling argument along the lines of Theorem~4.12.3 in \cite{grimmett01}.
}
Fisher's combined statistic will tend to be small when all stratum-level null hypotheses are true.
If any is false, then as the sample size increases, Fisher's combined statistic will tend to grow.

If, for all $\mathbf{\lambda}$ with $\sum_s \lambda_s = 1$, we can reject the conjunction
hypothesis at level $\alpha$ 
(i.e., if the minimum value of Fisher's combined statistic over all $\lambda$ is larger than the $1-\alpha$ quantile of the chi-square
distribution with $2S$~degrees of freedom), the audit can stop. 

If the audit is allowed to ``escalate'' in steps, increasing the sample size sequentially, then either the tests used in the separate strata have to be sequential tests, or multiplicity needs to be taken into account, for instance by adjusting the risk limit at each step.
Otherwise, the overall procedure can have a risk limit that is much larger than $\alpha$.
For examples of controlling for multiplicity when using non-sequential testing procedures in an RLA, see \cite{stark08a,stark09a}.

The stratum-level $P$-value $p_s(\lambda)$ could be a $P$-value for the hypothesis
$\omega_{w\ell,s} \ge \lambda_s V_{w\ell}$ from any test procedure. 
We assume, however, that $p_s$ is based on a one-sided test, and that the tests
for different values of $\lambda$ ``nest'' in the sense that if $a > b$,
then $p_s(a) > p_s(b)$.
This monotonicity is a reasonable requirement because the evidence that the overstatement
is greater than $a$ should be weaker than the evidence that the overstatement is greater than
$b$, if $a > b$.
In particular, this monotonicity holds for the tests proposed in Sections~\ref{sec:comparisonError}
and \ref{sec:ballotPollError}.

One could use a function other than Fisher's to combine the stratum-level $P$-values into a $P$-value for the conjunction hypothesis, provided it satisfies these properties (see \cite{pesarinSalmaso10}):
\begin{itemize}
  \item the function is non-increasing in each argument and symmetric with respect to rearrangements of the arguments
  \item the combining function attains its supremum when one of the arguments approaches zero
  \item for every level $\alpha$, the critical value of the combining function is finite and strictly smaller than the function's supremum.
\end{itemize}
For instance, one could use Liptak's function, $T = \sum_i \Phi^{-1}(1-p_i)$,
or Tippett's function, $T = \max_i (1-p_i)$.

Fisher's function is convenient for this application because the tests in different strata are independent, so the chi-squared distribution dominates the distribution of $\chi(\cdot)$ when the null hypothesis is true.
If tests in different strata were correlated, the null distribution of the combination function would need to be calibrated by simulation; some other combining function might have better properties than Fisher's \cite{pesarinSalmaso10}.

\subsection{Maximizing Fisher's combined $P$-value for $S=2$}
We now specialize to $S=2$ strata.
The set of $\mathbf{\lambda} = (\lambda_1, \lambda_2)$ such that $\sum_s \lambda_s = 1$ is then a one-dimensional family: if $\lambda_1 = \lambda$, then $\lambda_2 = 1-\lambda$.
For a given set of data, finding the maximum $P$-value over all $\mathbf{\lambda}$
is thus a one-dimensional optimization problem.
We provide two software solutions to the problem.

The first approach approximates the maximum via a grid search, refining the
grid once the maximum has been bracketed.
This is not guaranteed to find the global maximum exactly, although it can approximate 
the maximum as closely as one desires by refining the mesh, since the objective function is continuous.

The second, more rigorous approach uses bounds on Fisher's combining function $\chi$ for all $\lambda$. 
(A lower bound on $\chi$ implies an upper bound on the $P$-value: if, for all $\lambda$, the lower bound is 
larger than the $1-\alpha$ quantile of the chi-squared distribution with 4~degrees of freedom, the maximum $P$-value is no larger than $\alpha$.)

Some values of $\lambda$ can be ruled out \emph{a priori}, because (for instance) $\omega_{w\ell,s} \le V_{w\ell,s}+N_s$, where $N_s$ is the number of ballots cast in stratum $s$, and thus
\beq
   1 - \frac{V_{w\ell,2}+N_2}{V_{w\ell}} \le \lambda \le \frac{V_{w\ell,1}+N_1}{V_{w\ell}}.
\eeq
Let $\lambda_-$ and $\lambda_+$ be lower and upper bounds on $\lambda$.

Recall that $p_s(\cdot)$ are monotonically increasing functions, so, as a function of $\lambda$, $p_1(\lambda)$ increases monotonically and $p_2(1-\lambda)$ decreases monotonically.
Suppose $[a, b) \subset [\lambda_-, \lambda_+]$.
Then for all $\lambda \in [a, b)$, $-2\ln p_1(\lambda) \ge -2\ln p_1(b)$ and
$-2\ln p_2(1-\lambda) \ge -2\ln p_2(1-a)$.
Thus
\beq\label{eq:lowerbound}
   \chi(\lambda) = -2(\ln p_1(\lambda)+ \ln p_2(1-\lambda))
          \ge -2(\ln p_1(b) + \ln p_2(1-a)) \equiv \chi_-[a,b).
\eeq
This gives a lower bound for $\chi$ on the interval $[a, b)$; the corresponding 
upper bound is $\chi(\lambda) \le -2(\ln p_1(a) + \ln p_2(1-b)) \equiv \chi_+[a,b)$.
Partitioning $[\lambda_-, \lambda_+]$ into a collection of intervals $[a_k, a_{k+1})$
and finding $\chi_-[a_k, a_{k+1})$ and $\chi_+[a_k, a_{k+1})$ for each
yields piecewise-constant lower and upper bounds for $\chi(\lambda)$.

If, for all $\lambda \in [\lambda_-, \lambda_+]$, the lower bound on $\chi$
is larger than the $1-\alpha$ quantile of the chi-square distribution with 4~degrees of freedom,
the audit can stop.
On the other hand, if for some $\lambda \in [\lambda_-, \lambda_+]$, 
the upper bound is less than the $1-\alpha$ quantile of the chi-square distribution with 4~degrees of freedom, 
or if $\chi(a_k)$ is less than this quantile at any grid point $\{a_k\}$,
the sample size in one or both strata needs to increase.
If the lower bound is less than the $1-\alpha$ quantile on some interval,
but $\chi(a_k)$ is above this quantile at every grid point $\{a_k\}$, then one should improve the lower bound by refining the grid and/or by increasing the sample size in one or both strata.

\section{Auditing cross-jurisdictional contests}\label{sec:combiningMethods}
As mentioned above, stratified sampling can simplify audit logistics by allowing jurisdictions
to sample ballots independently of each other, or by allowing a single jurisdiction to sample 
independently from different collections of ballots (e.g., vote-by-mail versus cast in person).
SUITE allows stratified samples to be combined into an RLA of contests that include ballots from more
than one stratum.

We present an example where SUITE is helpful for a different reason: 
it enables an RLA to take
advantage of differences among voting systems to reduce audit sample sizes, which solves
a current problem in Colorado.

CRS~1-7-515 requires Colorado to conduct risk-limiting audits beginning in 2017.
The first risk-limiting election audits under this statute were conducted in November, 2017; the second were conducted in July, 2018.\footnote{%
 See \url{https://www.sos.state.co.us/pubs/elections/RLA/2017RLABackground.html}
}
Counties cannot audit contests that cross jurisdictional boundaries (\emph{cross-jurisdictional} contests, such as gubernatorial contests and most federal contests)
on their own: margins and risk limits apply to entire contests, not to the portion of a 
contest included in a county.
Colorado has not yet conducted an RLA of a cross-jurisdictional contest, although it has performed RLA-like procedures on individual jurisdictions' portions of some cross-jurisdictional contests.
To audit statewide elections and contests that cross county lines, Colorado will need to implement new approaches and make some changes to its auditing software, RLATool.

Colorado's voting systems are heterogeneous. 
Some counties (containing about 98\% of active voters, as of this writing) have 
voting systems that export cast vote records (CVRs) in a way that the paper ballot corresponding to each CVR can be identified uniquely and retrieved.
We call counties with such voting systems \emph{CVR counties}.
In CVR counties, auditors can manually check the accuracy of the voting system's interpretation of individual ballots.
In other counties (\emph{legacy} or \emph{no-CVR} counties) there is no way to check the accuracy
of the system's interpretation of voter intent for individual ballots.

Contests entirely contained in CVR counties can be audited using \emph{ballot-level comparison audits} \cite{lindemanStark12}, which compare CVRs to the auditors' interpretation of voter intent directly from paper ballots.
Ballot-level comparison audits are currently the most efficient approach to risk-limiting audits in that they require examining fewer ballots than other methods do, when the outcome of the contest under audit is in fact correct.
Contests involving no-CVR counties can be audited using \emph{ballot-polling audits} \cite{lindemanEtal12,lindemanStark12},  which generally require examining more ballots than ballot-level comparison audits to attain the same risk limit.

Colorado's challenge is to audit contests that include ballots cast in both CVR counties and no-CVR counties. There is no literature on how to combine ballot polling with ballot-level comparisons to audit 
cross-jurisdictional contests 
that include voters in CVR counties and voters in no-CVR counties.\footnote{%
  See~\cite{Rivest-2018-bayesian-tabulation-audits}
  for a different (Bayesian) approach to auditing contests that include both CVR counties
  and no-CVR counties. In general, Bayesian audits are not risk-limiting.
}

Colorado could simply revert to ballot-polling audits for cross-jurisdictional contests that include votes in no-CVR counties, but that would entail a loss of efficiency.
Alternatively, Colorado could use batch-level comparison audits, with single-ballot batches in CVR counties and larger batches in no-CVR counties.\footnote{%
Since so few ballots are cast in no-CVR counties, cruder approaches might work, for instance, pretending that no-CVR counties had CVRs, but treating any ballot sampled from a no-CVR county as if it had a 2-vote overstatement error. See \cite{banuelosStark12}.
}
The statistical theory for such audits has been worked out (see, e.g., \cite{stark08a,stark09c,stark09b,stark10d} and Section~\ref{sec:appendix-comparison}, below); indeed, this is the method that was used in several of California's pilot audits, including the audit in Orange County, California.
However, batch-level comparison audits were found to be less efficient than ballot-polling audits in these pilots \cite{CA_SOS_EAC}.

Moreover, to use batch-level comparison audits in Colorado would require major changes to RLATool, for reporting batch-level contest results prior to the audit, for drawing the sample, for reporting audit findings, and for determining when the audit can stop. 
The changes would include modifying data structures, data uploads, random sampling procedures, and the county user interface.
No-CVR counties would also have to revise their audit procedures.
Among other things, they would need to report vote subtotals
for physically identifiable groups of ballots before the audit starts.
No-CVR counties with voting systems that can only report subtotals by precinct
might have to make major changes to how they handle ballots, for instance, sorting all ballots by precinct.
These are large changes.

We show here that SUITE makes possible a ``hybrid'' RLA that keeps the advantages of ballot-level comparison audits in CVR counties but does not require major changes to how no-CVR counties audit, nor major changes to RLATool. 
The key is to use stratified sampling with two strata: ballots cast in CVR counties and those cast in no-CVR counties.

In order to use Equation~\ref{eq:fisher}, we must develop stratum-level tests for the overstatement error that are appropriate for the corresponding voting system.
Sections~\ref{sec:comparisonError} and~\ref{sec:ballotPollError} describe these tests for overstatement in the CVR and no-CVR strata, respectively.

\subsection{Comparison audits of overstatement quotas}
\label{sec:comparisonError}

To use comparison auditing in the approach to stratification described above requires extending previous work to test whether the overstatement error is greater than or equal to $\lambda_s V_{w\ell}$, rather than simply $V_{w\ell}$.
Appendix~\ref{sec:appendix-comparison} derives this generalization for arbitrary batch sizes, including batches consisting of one ballot.
The derivation considers only a single contest, but the 
MACRO test statistic \cite{stark09c,stark10d} automatically extends the result to 
auditing any number of contests simultaneously.
The derivation is for plurality contests, including ``vote-for-$k$'' plurality contests.
Majority and super-majority contests are a minor 
modification~\cite{stark08a}.\footnote{%
  So are some forms of preferential and approval voting, such as Borda count, and
  proportional representation contests, such as D'Hondt~\cite{starkTeague14}.
  For a derivation of ballot-level comparison risk-limiting audits for super-majority contests, 
  see \url{https://github.com/pbstark/S157F17/blob/master/audit.ipynb}. (Last visited 14~May 2018.)
  Changes for IRV/STV are more complicated.
}

\subsection{Ballot-polling audits of overstatement quotas}
\label{sec:ballotPollError}

To use the new stratification method with ballot polling requires a different approach than \cite{lindemanEtal12} took: their approach tests whether $w$ got a larger \emph{share} of the votes than $\ell$, but we need to test whether the margin \emph{in votes} in the stratum is greater than or equal to a threshold (namely, $\lambda_s V_{w\ell}$).
This introduces a nuisance parameter, the number of ballots with votes for either $w$ or $\ell$.
We address this by maximizing the probability ratio in Wald's Sequential Probability Ratio Test 
\cite{wald45} over all possible values of the nuisance parameter.
Appendix~\ref{sec:appendix-polling} develops the test.

\section{Numerical examples}\label{sec:examples}

Jupyter notebooks containing calculations for hybrid stratified audits intended to be relevant for Colorado are available at \url{https://www.github.com/pbstark/CORLA18}.

\texttt{hybrid-audit-example-1} contains two hypothetical examples. 
The first has $110,000$ cast ballots, of which 
9.1\% were in no-CVR counties. 
The \emph{diluted margin} (the margin in votes, divided by the total number of ballots cast) is $1.8\%$.
In 94\% of 10,000 simulations in which
the reported results were correct, drawing 700~ballots from the CVR stratum and 500~ballots 
from the no-CVR stratum (1,200 ballots in all)
allowed SUITE to confirm the outcome at 10\% risk.
For the remaining 6\%, further expansion of the audits would have been necessary.

If it were possible to conduct a ballot-level comparison audit for the entire contest, 
an RLA with risk limit 10\% could terminate after examining 263~ballots if it found no errors.
A ballot-polling audit of the entire contest would have been expected to examine about 14,000 ballots, more than 10\% of ballots cast.
The hybrid audit is less efficient than a ballot-level comparison audit, but far more efficient than a ballot-polling audit.

The second contest has 2~million cast ballots, of which 5\% were cast in no-CVR counties.
The diluted margin is about $20\%$.
The workload for SUITE at 5\% risk is quite low:
In 100\% of 10,000 simulations in which
the reported results were correct, auditing 43~ballots from the 
CVR stratum and 15~ballots from the no-CVR stratum
would have confirmed the outcome.
If it were possible to conduct a ballot-level comparison audit for the entire contest, 
an RLA at risk limit 5\% could terminate after examining 31~ballots if it found no errors.
The additional work for the hybrid stratified audit is disproportionately in the no-CVR counties.

A second notebook, \texttt{hybrid-audit-example-2}, illustrates the 
workflow for SUITE for an election with 2~million ballots cast.
The reported margin is just over $1\%$, but the reported winner
and reported loser are actually tied in both strata.  
The risk limit is 5\%.
For a sample of 500~ballots from the CVR stratum and 1000~ballots from the no-CVR stratum, 
the maximum combined $P$-value is over 25\%, so the audit cannot stop there.

A third notebook, \texttt{fisher\_combined\_pvalue}, illustrates the numerical methods used
to check whether the maximum combined $P$-value is below the risk limit.
It includes code for the tests in the two strata,
for the lower and upper bounds $\lambda_-$ and $\lambda_+$ for $\lambda$,
for evaluating Fisher's combining function on a grid,
and for computing bounds on the $P$-value via Equation~\ref{eq:lowerbound}.

\section{Discussion} \label{sec:discussion}

We present SUITE, a new class of procedures for RLAs based on stratified random sampling.
SUITE is agnostic about the capability of voting equipment in different strata, unlike
previous methods, which require batch-level comparisons in every stratum.
SUITE allows arbitrary tests to be used in different strata; if those tests are sequentially valid, then the overall RLA is sequential. 
(Otherwise, multiplicity adjustments might be needed if one wants an audit that escalates in stages.
See~\cite{stark08a,stark09a} for two approaches.)

Like other RLA methods, SUITE poses auditing as a hypothesis test.
The null hypothesis is a union over all partitions of outcome-changing error across strata.
The hypothesis is rejected if the maximum $P$-value over all such partitions is sufficiently
small.
Each possible partition yields an intersection hypothesis, tested by
combining $P$-values from different strata using Fisher's combining function (or a suitable
replacement). 

Among other things, the new approach solves a current problem in Colorado:
how to conduct RLAs of contests that cross jurisdictional lines, such as statewide 
contests and many federal contests.

We give numerical examples in Jupyter notebooks
that can be modified to estimate the workload for different contest sizes, margins, and risk limits.
In our numerical experiments, the new method requires auditing far fewer ballots than previous approaches would. 

\appendix
\section{Comparison tests for an overstatement quota}\label{sec:appendix-comparison}
\subsection{Notation}
\begin{itemize}
    \item  $\mathcal{W}$: the set of reported winners of the contest
    \item  $\mathcal{L}$: the set of reported losers of the contest
    \item  $N_s$ ballots were cast in stratum $s$. (The contest might not appear on all $N_s$ ballots.)
    \item  $P$ ``batches'' of ballots are in stratum $s$. A batch contains one or more ballots. Every ballot in stratum $s$ is in exactly one batch.
    \item  $n_p$: number of ballots in batch $p$. $N_s = \sum_{p=1}^P n_p$.
    \item  $v_{pi} \in \{0, 1\}$: reported votes for candidate $i$ in batch $p$
    \item  $a_{pi} \in \{0, 1\}$: actual votes for candidate $i$ in batch $p$. 
                  If the contest does not appear on any ballot in batch $p$, then $a_{pi} = 0$.
                  
    \item  $V_{w\ell,s} \equiv \sum_{p=1}^P (v_{pw} - v_{p\ell})$: 
Reported margin in stratum $s$ of reported winner $w \in \mathcal{W}$ over reported loser 
$\ell \in \mathcal{L}$, in votes.

    \item  $V_{w\ell}$: 
overall reported margin in votes of reported winner $w \in \mathcal{W}$ over reported loser 
$\ell \in \mathcal{L}$ for the entire contest (not just stratum $s$)

    \item  $V \equiv \min_{w \in \mathcal{W}, \ell \in \mathcal{L}} V_{w \ell}$: 
smallest reported overall margin in votes between any reported winner and reported loser

    \item  $A_{w\ell,s} \equiv \sum_{p=1}^P (a_{pw} - a_{p\ell})$: 
actual margin in votes in the stratum of reported winner $w \in \mathcal{W}$ over reported loser 
$\ell \in \mathcal{L}$

    \item  $A_{w\ell}$: 
actual margin in votes of reported winner $w \in \mathcal{W}$ over reported loser 
$\ell \in \mathcal{L}$ for the entire contest (not just in stratum $s$)

\end{itemize}

\subsection{Reduction to maximum relative overstatement}
If the contest is entirely contained in stratum $s$, then
the reported winners of the contest are the actual winners if
$$ 
   \min_{w \in \mathcal{W}, \ell \in \mathcal{L}} A_{w\ell,s} > 0.
$$
Here, we address the case that the contest may include a portion outside the stratum.
To combine independent samples in different strata, it is convenient
to be able to test whether the net overstatement error in a stratum is greater than or equal to a given threshold.

Instead of testing that condition directly, we will test a condition that is sufficient 
but not necessary for the inequality to hold, to get a computationally simple test that
is still conservative (i.e., the level is not larger than its nominal value).

For every winner, loser pair $(w, \ell)$, we want to test
whether the overstatement error is greater than or equal to some threshold, generally
one tied to the reported margin between $w$ and $\ell$.
For instance, for a hybrid stratified audit, we set the threshold to be
$\lambda_s V_{w\ell}$.

We want to test whether
$$
   \sum_{p=1}^P (v_{pw}-a_{pw} - v_{p\ell} + a_{p\ell})/V_{w\ell} \ge \lambda_s.
$$
The maximum of sums is not larger than the sum of the maxima; that is,
$$
\max_{w \in \mathcal{W},  \ell \in \mathcal{L}}
   \sum_{p=1}^P (v_{pw}-a_{pw} - v_{p\ell} + a_{p\ell})/V_{w\ell}
   \le
  \sum_{p=1}^P  \max_{w \in \mathcal{W},  \ell \in \mathcal{L}} 
  (v_{pw}-a_{pw} - v_{p\ell} + a_{p\ell})/V_{w\ell}.
$$

Define 
$$
  e_p \equiv \max_{w \in \mathcal{W} \ell \in \mathcal{L}} 
     (v_{pw}-a_{pw} - v_{p\ell} + a_{p\ell})/V_{w\ell}.
$$
Then no reported margin is overstated by a fraction $\lambda_s$ or more
if 
$$ 
  E \equiv \sum_{p=1}^P e_p < \lambda_s.
$$

Thus if we can reject the hypothesis $E \ge \lambda_s$, we can conclude that
no pairwise margin was overstated by as much as a fraction $\lambda_s$.

Testing whether $E \ge \lambda_s$ would require a very large sample if we knew nothing at
all about $e_p$ without auditing batch $p$: a single large value of $e_p$ could make
$E$ arbitrarily large.
But there is an \emph{a priori} upper bound for $e_p$.
Whatever the reported votes $v_{pi}$ are in batch~$p$, we can find the
potential values of the actual votes $a_{pi}$ that would make the
error $e_p$ largest, because $a_{pi}$ must be between 0 and $n_p$,
the number of ballots in batch~$p$:
$$
    \frac{v_{pw}-a_{pw} - v_{p\ell} + a_{p\ell}}{V_{w\ell}} \le 
    \frac{v_{pw}- 0 - v_{p\ell} + n_p}{V_{w\ell}}.
$$
Hence,
\begin{equation} \label{eq:uDef}
    e_p \le \max_{w \in \mathcal{W}, \ell \in \mathcal{L}} 
    \frac{v_{pw} - v_{p\ell} + n_p}{V_{w\ell}} \equiv u_p.
\end{equation}

Knowing that $e_p \le u_p$ might let us conclude reliably that $E < \lambda_s$
by examining only a small number of batches---depending on the 
values $\{ u_p\}_{p=1}^P$ and on the values of $\{e_p\}$ for the audited batches.

To make inferences about $E$, it is helpful to work with the \emph{taint} 
$t_p \equiv \frac{e_p}{u_p} \le 1$.
Define $U \equiv \sum_{p=1}^P u_p$.
Suppose we draw batches at random with replacement, with probability $u_p/U$
of drawing batch $p$ in each draw, $p = 1, \ldots, P$.
(Since $u_p \ge 0$, these are all positive numbers, and they sum to 1,
so they define a probability distribution on the $P$ batches.)

Let $T_j$ be the value of $t_p$ for the batch $p$ selected in the $j$th draw.
Then $\{T_j\}_{j=1}^n$ are IID, $\mathbb{P} \{T_j \le 1\} = 1$, and
$$
  \mathbb{E} T_1 = \sum_{p=1}^P \frac{u_p}{U} t_p =
  \frac{1}{U}\sum_{p=1}^P u_p \frac{e_p}{u_p} = 
  \frac{1}{U} \sum_{p=1}^P e_p = E/U.
$$
Thus $E = U \mathbb{E} T_1$. 
So, if we have strong evidence that
$\mathbb{E} T_1 < \lambda_s/U$, we have
strong evidence that $E < \lambda_s$.

This approach can be simplified even further by noting that $u_p$ has
a simple upper bound that does not depend on $v_{pi}$.
At worst, the reported result for batch $p$ shows $n_p$ votes for the 
``least-winning'' apparent winner of the contest with the smallest margin, 
but a hand interpretation would show that all $n_p$ ballots in the batch 
had votes for the runner-up in that contest.
Since 
$V_{w\ell} \ge V\equiv \min_{w \in \mathcal{W}, \ell \in \mathcal{L}} V_{w \ell}$ and $0 \le v_{pi} \le n_p$,
$$ 
    u_p =  \max_{w \in \mathcal{W}, \ell \in \mathcal{L}} 
    \frac{v_{pw} - v_{p\ell} + n_p}{V_{w\ell}}
    \le  \max_{w \in \mathcal{W}, \ell \in \mathcal{L}} 
    \frac{n_p - 0 + n_p}{V_{w\ell}}
    \le \frac{2n_p}{V}.
$$
Thus if we use $2n_p/V$ in lieu of $u_p$, we still get conservative results.
(We also need to re-define $U$ to be the sum of those upper bounds.)
An intermediate, still conservative approach would be to use this upper bound for
batches that consist of a single ballot, but use the sharper bound (\ref{eq:uDef})
when $n_p > 1$.
Regardless, for the new definition of $u_p$ and $U$,
$\{T_j\}_{j=1}^n$ are IID, $\mathbb{P} \{T_j \le 1\} = 1$,
and
$$
  \mathbb{E} T_1 = \sum_{p=1}^P \frac{u_p}{U} t_p =
  \frac{1}{U}\sum_{p=1}^P u_p \frac{e_p}{u_p} = 
  \frac{1}{U} \sum_{p=1}^P e_p = E/U.
$$

So, if we have evidence that $\mathbb{E} T_1 < \lambda_s/U$, we have evidence that 
$E < \lambda_s$.

\subsection{Testing $\mathbb{E} T_1 \ge \lambda_s/U$}

A variety of methods are available to test whether $\mathbb{E} T_1 < \lambda_s/U$.
One particularly elegant sequential method is based on Wald's Sequential Probability
Ratio Test (SPRT) \cite{wald45}.
Harold Kaplan pointed out this method on a website that no longer exists.
A derivation of this \emph{Kaplan-Wald} method is in Appendix A of~\cite{starkTeague14};
to apply the method here, take $t = \lambda_s$ in their equation~18.
A different sequential method, the \emph{Kaplan-Markov} method (also due to Harold Kaplan), 
is given in~\cite{stark09b}.

\section{Ballot-polling tests for an overstatement quota}\label{sec:appendix-polling}
In this section, we derive a ballot-polling test of the hypothesis that the margin (in votes) in a single stratum is greater than or equal to a threshold $c$. 

\subsection{Wald's SPRT with a nuisance parameter}

Consider a single stratum $s$ containing $N_s$ ballots, of which 
$N_{w,s}$ have a vote for $w$ but not for $\ell$, $N_{\ell,s}$ have a vote for $\ell$ but not for $w$, and $N_{u,s} = N_s - N_{w,s} - N_{\ell,s}$ have votes for both $w$ and $\ell$ or neither $w$ nor $\ell$, including undervotes and invalid ballots.
Ballots are drawn sequentially without replacement, with equal probability of selecting each as-yet-unselected ballot in each draw.

We want to test the compound hypothesis that $N_{w,s} - N_{\ell,s} \le c$ against the alternative that $N_{w,s} = V_{w,s}$, $N_{\ell,s} = V_{\ell,s}$, and $N_{u,s} = V_{u,s}$, with $V_{w,s} - V_{\ell,s} > c$.

The values $V_{w,s}$, $V_{\ell,s}$, and $V_{u,s}$ are the reported results for stratum $s$ 
(or values related to those reported results; see \cite{lindemanEtal12}). 
In this problem, $N_{u,s}$ (equivalently, $N_{w,s} + N_{\ell,s}$) is a nuisance parameter: we care about $ N_{w,s} - N_{\ell,s}$.

Let $X_k$ be $w$, $\ell$, or $u$ according to whether the ballot selected on the $k$th draw 
shows a vote for $w$ but not $\ell$, $\ell$ but not $w$, or something else.
Let $W_n \equiv \sum_{k=1}^n 1_{X_k = w}$; and define $L_n$ and $U_n$ analogously.

The probability of a given data sequence $X_1, \ldots, X_n$ under the alternative hypothesis is
$$
    \frac{\prod_{i=0}^{W_n-1} (V_{w,s}-i) \; 
             \prod_{i=0}^{L_n-1} (V_{\ell,s}-i) \;
             \prod_{i=0}^{U_n-1} (V_{u,s}-i)}
            {\prod_{i=0}^{n-1} (N_s-i)}.
$$
If $L_n \ge W_n - cn/N_s$, the data obviously do not provide evidence against the null, so we suppose that
$L_n < W_n - cn/N_s$, in which case, the element of the null that will maximize the probability of the observed data has $N_{w,s} - c = N_{\ell,s}$.
Under the null hypothesis, the probability of $X_1, \ldots, X_n$ is
$$
    \frac{ \prod_{i=0}^{W_n-1} (N_{w,s}-i) \;
             \prod_{i=0}^{L_n-1}(N_{w,s}-c - i)
             \prod_{i=0}^{U_n-1} (N_{u,s}-i)}
             {\prod_{i=0}^n (N_s-i)},
$$
for some value $N_{w,s}$ and the corresponding $N_{u,s} = N_s - 2N_{w,s}+c$.
How large can that probability be under the null? 
The probability under the null is maximized by any integer 
$x \in \{ \max(W_n, L_n+c), \ldots, (N-U_n)/2 \}$ that maximizes 
$$
\prod_{i=0}^{W_n-1} (x-i) \; \prod_{i=0}^{L_n-1} (x-c-i) \; \prod_{i=0}^{U_n-1} (N_s-2x+c - i).
$$
The logarithm is monotonic, so any maximizer $x^*$ also maximizes
$$ f(x) = \sum_{i=0}^{W_n-1} \ln (x-i) + \sum_{i=0}^{L_n-1} \ln (x-c-i) + \sum_{i=0}^{U_n-1} \ln (N_s-2x+ c - i).$$
The first two terms on the right increase monotonically with $x$ and the last term decreases monotonically with $x$.
This yields bounds without having to evaluate $f$ everywhere.
Suppose $y < z$. 
Then for all integer $x$ between $y$ and $z$, 
$$ f(x) \le \sum_{i=0}^{W_n-1} \ln (z-i) + \sum_{i=0}^{L_n-1} \ln (z-c-i) + \sum_{i=0}^{U_n-1} \ln (N_s-2y+c-i).$$
The optimization problem can be solved using a branch and bound approach.
For instance, start by evaluating 
$$
   f^+(x) \equiv \sum_{i=0}^{W_n-1} \ln (x-i) + \sum_{i=0}^{L_n-1} \ln (x-c-i)
$$
and
$$
  f^-(x) \equiv \sum_{i=0}^{U_n-1} \ln (N_s-2x+c-i)
$$
at $\max(W_n, L_n+c)$, $(N_s-U_n)/2$, and their midpoint,
 to get the values of $f = f^+ + f^-$ at those three points, along with 
upper bounds on $f$ on the ranges between them.
At stage $j$, we have evaluated $f$, $f^+$, and $f^-$ at $j$ points $x_1 < x_2 < \ldots < x_j$, and 
we have upper bounds on $f$ on the $j-1$ ranges $R_m = \{x_m, x_m+1, \ldots, x_{m+1}\}$
between those points.
Let $U_m$ be the upper bound on $f(x)$ for $x \in R_m$.
Suppose that for some $h$, $f(x_h) = \max_{m=1}^j U_m$.
Then $x^* = x_h$ is a global maximizer of $f$.
If there is some $U_m > \max_i f(x_i)$, then subdivide the range with the largest $U_m$,
calculate $f$, $f^+$, and $f^-$ at the new point, and repeat.
This algorithm must terminate by identifying a global maximizer $x^*$ after a finite number of steps.

A conservative $P$-value for the null hypothesis after $n$ items have been drawn is thus
$$
   P_n =  \frac{\prod_{i=0}^{W_n-1} (x^*-i) \; \prod_{i=0}^{L_n-1}  (x^*-c-i) \; \prod_{i=0}^{U_n-1} (N_s-2x^*+c-i)}{\prod_{i=0}^{W_n-1}(V_{w,s}-i) \; \prod_{i=0}^{L_n-1} (V_{\ell,s}-i) \; \prod_{i=0}^{U_n-1} (V_{u,s}-i)}.
$$
Because the test is built on Wald's SPRT,  the sample can expand sequentially and 
(if the null hypothesis is true) the chance that $P_n < p$ is never larger than $p$.
That is, $\Pr \{ \inf_n P_n < p \} \le p$ if the null is true.

A Jupyter notebook implementing this approach is given in \url{https://github.com/pbstark/CORLA18}.

\bibliography{./pbsBib}

\end{document}